\documentclass{iau} 
\usepackage{graphicx}
\usepackage{orcidlink}

\title[The Magnetar Connection] 
{The Magnetar Connection}

\author[Chowhan, Konar \& Banik]   
{Tanmay Tushar Chowhan$^1$\orcidlink{0000-0003-0849-4773}, Sushan Konar$^2$\orcidlink{0000-0003-2231-6658}, Sarmistha Banik$^1$\orcidlink{0000-0003-0221-3651}}

\affiliation{$^1$Dept. of Physics, BITS Pilani, Hyderabad Campus, \\ Shameerpet Mandal,
Hyderabad-500078, India. \\[\affilskip]
$^2$NCRA-TIFR, Pune 411007, India}

\pubyear{2021}
\volume{363}  
\jname{Neutron Star Astrophysics at the Crossroads: Magnetars and the Multimessenger Revolution}
\editors{Prof. E. Troja \& Prof M. Baring}
\begin{document}

\maketitle

\begin{abstract}
We investigate the combined evolution  of the dipolar surface magnetic
field (B$_{s}$) and the spin-period (P$_s$) of known magnetars and high  
magnetic field (B$_s$~$ \gtrsim  10^{13}$~G) radio pulsars. We study the 
long term behaviour of these objects assuming a simple ohmic dissipation 
of the magnetic field. Identifying the regions (in the P$_s$-B$_s$ plane) 
in which these neutron stars would likely move into, before crossing the 
death-line to enter the pulsar graveyard, we comment upon the possible 
connection between the magnetars and other classes of neutron stars.

\end{abstract}

\noindent
More  than fifty  years have  passed  since the  discovery of  neutron
stars. In  this period,  $\sim$3500 neutron  stars, belonging  to many
distinct observational classes have been observed. This has led to an
important  direction of  neutron star  research, in  trying to  find a
unification  scheme  (evolutionary   or  otherwise)  connecting  these
different observational classes. In particular, the connection between
different types of isolated neutron  stars (for example, the Magnetars,
RRATs,  CCOs, XDINS,  and ordinary  radio pulsars),  via the  route of
magnetic  field evolution,  has received  serious attention  in recent
years \cite[(Kaspi 2011)]{Kaspi11}.
Therefore,  it is  worthwhile  to study  the  time-trajectory of  such
isolated neutron stars in the spin-period - surface magnetic field (P$_s$-B$_s$) plane to understand possible
connections between different observational classes. With this aim, we
consider the  evolution of the  magnetars and the high  magnetic field
(B$_s$~$ \gtrsim 10^{13}$~G) radio pulsars. \\

\begin{figure}
  \begin{center}
   \includegraphics[width=7.5cm]{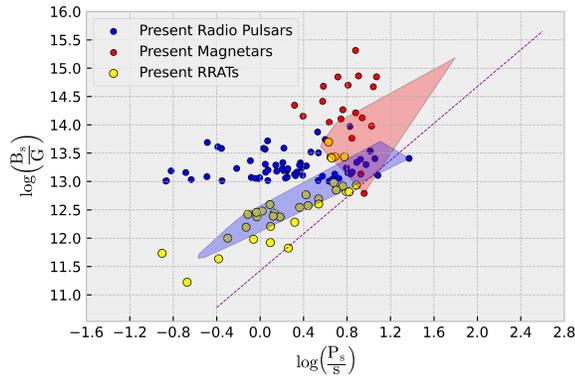} 
  \end{center}
   \caption{The current locations of  the magnetars, the high magnetic
     field radio pulsars alongwith the  region where they evolve into,
     in the P$_s$-B$_s$ plane. A number  of RRATs have also been shown
     for comparison.}
   \label{fig1}
\end{figure}

\noindent
The  exact  nature  of  the  magnetic field of neutron stars is  not  yet  clearly
understood.  In this  work,  we  assume a  purely  crustal field  (the
currents supporting  the field  are entirely  confined to  the crust)
evolving due to simple ohmic dissipation given by -
\setcounter{equation}{0}
\renewcommand{\theequation}{\arabic{equation}}

\begin{equation}
  \frac{\partial {\bf B}}{\partial t}
  = - \frac{c^2}{4\pi} \nabla \times (\frac{1}{\sigma} \nabla \times {\bf B}) \,.
\label{eqn01}
\end{equation}
We  use the  formalism for  field evolution  developed by  \cite[Konar
  (1997)]{Konar97}.  The effectiveness of  ohmic dissipation for field
evolution  has  recently  been  reiterated  by  \cite[Ertan  \&  Alpar
  (2021)]{Ertan21}  by  providing  an   explanation  for  the  minimum
spin-period seen in the millisecond pulsars.
The free parameters of the  model, entering the above equation through
the electrical conductivity ($\sigma$), are - a) the density ($\rho_c$)
at which  the currents are  concentrated, and b) the  impurity content
($Q$) of the  crustal material.  Since there is no  way of knowing the
exact parameter  values applicable for  a particular neutron  star, we
calculate the trajectory of each  object assuming the following ranges
for these parameters - a) $\rho_c \sim 10^{13} - 10^{11}$~gm.cm$^{-3}$,
b) $Q \sim 0.0 - 0.05$. \\

\noindent
The  fundamental  measured  quantities  of  a  neutron  star  are  its
spin-period (P$_s$)  and the period derivative  ($\dot{\rm P}_s$). For
the  rotation-powered   pulsars  (as   well  as  the   Magnetars)  the
large-scale dipolar  magnetic field, at  the surface, is  derived from
the following relation (\cite[Manchester \& Taylor 1977]{Manchester77}) -
\begin{equation}
\rm B_{s} = 3.2 \times 10^{19}
      \left(\frac{{\rm P}_{s}}{s}\right)^{1/2}
      \left(\frac{\dot{\rm P}_{s}}{s s^{-1}}\right)^{1/2}G\,. 
\label{eqn02}
\end{equation}
assuming that the magnetic dipole  radiation is solely responsible for
the spin-down of the pulsar. \\

\noindent
We track the evolutionary trajectories of - a) 19 Magnetars, and b) 65
radio  pulsars (with  B$_s  \gtrsim 10^{13}$~G,  beginning with  their
current values of  P$_s$ and B$_s$.  [The present values  of P$_s$ and
  B$_s$  for  both the  radio  pulsars  and  the Magnetars  have  been
  obtained  from  the ATNF  pulsar  catalog.]   Thereafter, with  each
incremental  change  in B$_s$,  we  dynamically  calculate P$_s$.  The
Magnetars are evolved for $2 \times  10^5$ years and the radio pulsars
for $5  \times 10^5$~years.  The timescales are  chosen such  that the
trajectories are arrested before  they reach the death-line \cite[(Chen
  \& Ruderman, 1993)]{Chen93} shown as a purple dashed line in Fig.[1]. \\

\noindent
Our results  are shown in  Fig.[1]. The shaded  areas in red  and blue
correspond to  the regions where  the Magnetars and the  radio pulsars
are found  at the  end of  the evolution.  Interestingly, most  of the
currently  known RRATs  are located  within this  region. In  a recent
work,  we have  shown  that the  population  of RRATs  do  not have  a
positive correlation  (statistically speaking) with the  population of
nulling pulsars, as is usually assumed \cite[(Abhishek et al. 2022)]{Abhishek22}. The current results, connecting the RRATs to magnetars and high-magnetic field radio pulsars through an evolutionary pathway, appear to corroborate this statistical conclusion.


\begin{thebibliography}{}

\bibitem[Abhishek]{Abhishek22}
  {Abhishek et al., } 2022
  \textit{astro-ph}, 2201.00295 

\bibitem[Chen]{Chen93}
  {Chen K., Ruderman M.,} 1993,
  \textit{ApJ}, 402, 264

\bibitem[Ertan]{Ertan21}
  {Ertan, Ü \& Alpar, M. A., } 2021
  \textit{MNRAS}, 505, L112

\bibitem[Kaspi]{Kaspi11}
  {Kaspi V. M.,} 2011,
  \textit{ApJ}, 741, L13A

\bibitem[Konar 1997]{Konar97}
  {Konar, S.,} 1997
  \textit{PhD thesis}, Indian Institute of Science, Bangalore 

\bibitem[Manchester]{Manchester77}
  {Manchester, R. N., Taylor, J. H.} 1977,
  \textit{Pulsars, W. H. Freeman, San Francisco}, p.36
  
\bibitem[Manchester]{Manchester93}
  {Manchester, R. N., Hobbs, G.B., Teoh, A. \& Hobbs, M.,} 1993-2006 (2005),
  \textit{AJ}, 129

\end{thebibliography}
\end{document}